\documentclass[a4paper]{aipproc}

\layoutstyle{6x9}
  
\begin{document}

\title {Quark degrees of freedom in hadronic systems}


\author{V. Vento}{
  address={Departamento de F\'{\i}sica Te\'orica,
Universidad de Valencia, 46100 Burjassot (Valencia), Spain},
  email={vicente.vento@uv.es},
  thanks={Supported in part by PB97-1127}
}

% \copyrightholder{Acoustical Scociety of America}
\copyrightyear  {2001}

\begin{abstract} 
The role of models in Quantum Chromodynamics is to produce
simple physical pictures that connect the phenomenological
regularities with the underlying structure. The static
properties of hadrons have provided experimental input to
define a variety of very successful Quark Models. We discuss
applications of some of the most widely used of these models 
to the high energy regime, a scenario for which they were
not proposed. The initial assumption underlying our
presentation will be that gluon and sea bremsstrahlung
connect the constituent quark momentum distributions with
the  partonic structure functions. The results obtained are
encouraging but lead to the necessity of  more complex
structures at the hadronic scale. This initial
hypothesis may be relaxed by introducing some non
perturbative model for the constituent quarks. Within
this scheme we will discuss some relevant problems
in nucleon structure as seen in high energy experiments.
\end{abstract}

\date{\today}

\maketitle

\section{Introduction}

The constituent quark, one of the most fruitful concepts in
20th century physics, was proposed to explain the structure
of the large number of hadrons being discovered in the
sixties \cite{Gell-Mann}. Soon thereafter deep inelastic
scattering of leptons off protons was explained in terms of
pointlike constituents named partons \cite{Feynman}. Thus
already at a very early stage of the study of  hadron
structure the need to connect the laboratory description,
based on constituent quarks, and the light cone
description, based on partons, arose as the way to
understand phenomena at different scales. Sum rules and
current algebra were very powerful tools to establish a
conceptual link between the two descriptions.

The birth of Quantum Chromodynamics ($QCD$) and the proof
that it is asymptotically free set the framework for an
understanding of deep inelastic phenomena beyond the parton
model \cite{QCD}. However, the perturbative approach to
$QCD$ does not provide absolute values for the observables,
it just gives their variation with momentum in terms of
some unknown non perturbative matrix element. On order for
the description based on  the Operator Product Expansion (OPE)
and $QCD$ evolution  to be  predictive, 
these matrix elements have to be eliminated by comparing several
processes or by the input of  experimental
data. Therefore the perturbative scheme is used, most of
the time, to relate experiments at different momentum
scales.

The phenomenological analysis proceeds by finding a
parametrization which is appropriate at a sufficiently
large momentum $Q^2_0$, where one expects perturbation
theory to be fully applicable, and then $QCD$ evolution
techniques determine the parton distributions at a
higher $Q^2$. As an example we show the  parametrization
due to Martin Sterling and Roberts ($Q^2_0 = 4 \;GeV ^2$)
\cite{RAL94}: 

\begin{eqnarray} 
x u_v & = & 2.26 x^{0.559}(1-0.54\sqrt{x} +
4.65x)(1-x)^{3.96}\nonumber \\ 
x d_v & = & 0.279 x^{0.335}(1+6.80\sqrt{x} +
1.93x)(1-x)^{4.46} \nonumber \\ 
x S & = & 0.956 x^{-0.17}(1-2.55\sqrt{x} +
11.2x)(1-x)^{9.63} \nonumber \\
x g & = &1.94 x^{-0.17}(1-1.90\sqrt{x} + 4.07x)(1-x)^{5.33}
\end{eqnarray} 

This parametrization incorporates the flavor and momentum
sum rules. The distributions are defined in the $\overline
{MS}$ renormalization and factorization schemes and the
$QCD$ scale paramenter $\Lambda$ is found to be $0.231\;
GeV$. With this fit a large body of data is
reasonably described. However this parametrization is
purely phenomenological with little theoretical input. 

The work of Gl\"uck, Reya and Vogt \cite{GRV} has shown
that the high energy parton distributions when evolved to a
low scale appear to indicate that a valence picture of
hadron structure arises. This idea was  suggested a long
time ago by Parisi and Petronzio \cite{Parisi}, who assumed
that there exists a low energy scale  $\mu_0^2$ such that
the glue and sea are absent, i.e., the long range part of
the  interaction (confinement)  is composed  in the
$P_\infty$ frame  of only three quarks. If one turns on 
the short range part of  the interaction (perturbative
$QCD$),  using the renormalization  group one introduces
gluons and the  sea. 

The constituent quark concept embedded in a $QCD$
framework, leads to models that are able to reproduce in an
extraordinary way the low energy properties with very few
parameters \cite{models}. The goal was to use them as
substitutes for $QCD$ at low energies. The needed
ingredient was provided by Jaffe and Ross \cite{Jaffe}.
According to these authors the quark model calculation of
matrix elements give their values at a hadronic scale
$\mu_0^2$ and for all larger $Q^2$ their coefficient
functions evolve as in perturbative $QCD$. 

We have developed a formalism, for potential quark models,
based on these ideas which connects the parton
distributions with the momentum distributions of the model 
\cite{Traini}. A analogous procedure may be derived for bag
models by using the bag model limit of light cone matrix
elements \cite{betz}. The low energy scale $\mu^2_0$ is
determined by evolving downward from the high energy data
the second moment of the  valence quark distribution until
it reaches the value given by the quark model describing
the  hadronic behavior. The model provides the matrix 
elements of the needed twist operators characterizing
observables at the high energy scale and their values are
ascribed to this hadronic scale. Then, they are evolved to
high momentum transfers, where comparison  with experiments
takes place, using perturbative $QCD$.

This approach describes successfully the gross features of
the DIS results \cite{Traini}. In order to produce more
quantitative fits different mechanisms have been proposed:
{\it valence} gluons, sea quarks and antiquarks,
relativistic kinematics, etc...  We will show   that some
of these mechanisms appear naturally if we endow the
constituent  quarks with structure  following the work of
Altarelli et al.\cite{acmp}. In our scheme {\it
constituent} quarks are complex  objects, made up of
point-like partons ({\it current} quarks (antiquarks) and
gluons), interacting by a residual interaction described by
a quark model \cite{scopetta}.  The hadron structure
functions are obtained as convolutions of the constituent 
quark wave function with the constituent quark structure
functions.

Our aim here is neither technical nor bibliographical. We
will simply guide the reader to the literature by
discussing the physics behind the various formalisms. In
the referred literature he will find a complete account of
the needed references and technicalities, so that he may be
able to reconstruct  the calculations presented in detail.
We will elaborate on the theoretical framework, discuss some
of the main results and explore future directions.

\section{Constituent quarks and partons}
Constituent quark models have been designed to describe the
static properties of hadrons and therefore aimed at
modeling the non-perturbative aspects of $QCD$. They are
in general very successful in their performance. We 
discuss a formalism which uses them to describe high energy
data, whose basis lies on the following reasoning. $QCD$
perturbation theory is non predictive. The renormalization
group relates different momentum scales.  Experimental 
input is required to avoid the unknown non-perturbative
properties of the theory.  Our formalism  substitutes the
experimental input by model physics. In this way we define
a predictive scheme, whose appeal lies in the relation it
establishes between physics at very different scales and
whose weakness is its model dependence.

\subsection{Parton distributions from quark models}
The basic idea in our approach arises from rephrasing the
OPE which states that, 

\begin{equation}  
F_i^n(Q^2) = M^n_{ij}F_j^n(Q_0^2), 
\end{equation}   
i.e., the nth moment of  structure functions at one scale
are related by means of perturbatively calculable
transformation matrices to the same moments at another
scale \cite{QCD}. If $Q_0^2$ is taken to be a low
scale,which we have named hadronic scale, the F functions
become highly non perturbative matrix elements in general.
We substitute the matrix elements at the hadronic scale by
the matrix elements calculated in the chosen model. In
particular we are able to relate the valence quark
distribution functions with the appropriate momentum
distributions in the corresponding baryonic state $n_q^a$,
i.e. with the hadronic wave functions in the model, 

\begin{equation} xq_V^a(x) = \frac{1}{(1-x)^2}\int
d^3p \;n_q^a(\vec{p})\; \delta (\frac{x}{1-x} -
\frac{p_+}{M}) 
\label{traini}
\end{equation} 
where $a$ represents the diverse degrees of freedom
(unpolarized, $\uparrow$, $\downarrow$, $\ldots$), $\vec{p}$
the momentum of the constituent, $p_+ =
p_0 - p_z$, $x$ is the Bjorken variable and $M$ the mass of
the baryonic state.

In this way we have studied polarized and unpolarized
structure function, transversity distributions and angular
momentum distributions with various models
\cite{Traini,scopetta1,scopetta2}. The results of our
calculations show that these models, with the parameters
fixed by low energy properties are able to provide a
qualitative description of the data and therefore the scheme
becomes predictive. They are however too naive and new
ingredients, not seen by low energy probes, have to be
incorporated.

\begin{figure}
\caption{We show the unpolarized parton distribution $xu_v$:
i) for a quark model \cite{bil} at the hadronic scale (dot
dashed); ii)
for the same model within the convolution approach at the
hadronic scale (long
dashed); iii) evolved (NLO) to the scale of the data at $10
GeV^2$ for the model in i) (dashed); iv) evolved for the
convolution approach  of ii) (full
curve) to the scale of the data; v) as guide line through
the data
(dotted) \cite{dataup}.}
\rotatebox{270}{\includegraphics[height=.5\textheight]{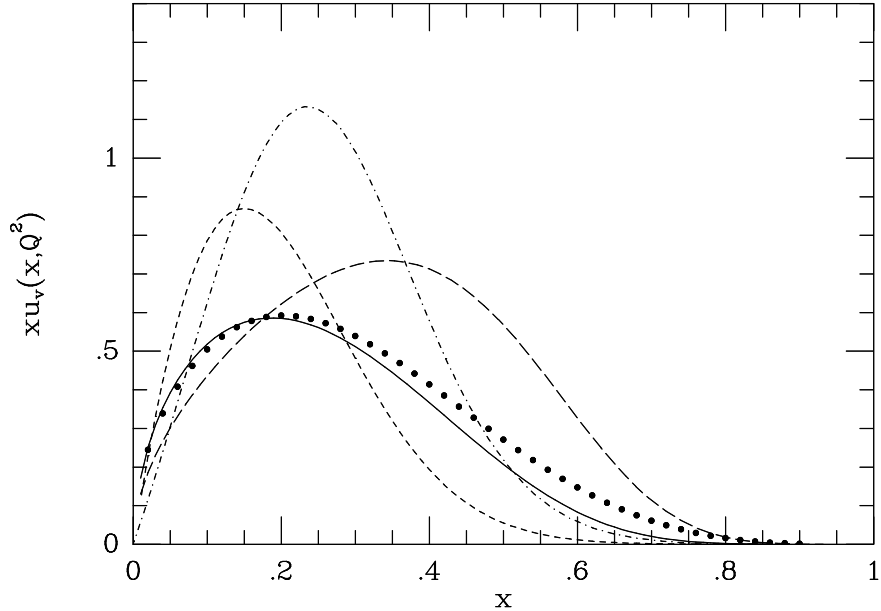}}
\end{figure}

\subsection{Applications}

We comment on some of the calculations performed by
stressing only the main results. We refer the reader to the
figures and discussions in the given references for a
complete account.

{\bf 1) Parton distributions \cite{Traini}}

We have analyzed in this formalism the polarized and
unpolarized experimental results and have shown that
well-known Quark Models lead to a qualitative description
of the data. The relevant features are: the original model
distributions, which are vastly different from the data,
evolve, via the Renormalization Group, towards them;  sea
quarks and gluons, initially absent, are generated  by
bremsstrahlung. In Fig. 1 one can see how the initial large
quark model distribution at the hadronic scale approaches
the data by evolution. The momentum lost by the valence
quarks goes into the other components. In Fig. 2 we show the
gluonic component obtained by evolution in a scheme to be
specified later.

\begin{figure} 
\caption{We show the gluon distribution $xg(x,Q^2)$ at 
$Q^2 = 10 GeV$ obtained with the ACMP approach for two 
different models of hadron structure \cite{bil,ik}. The data
are those of ref. \cite{dataup}.} 
\rotatebox{270}{\includegraphics[height=.5\textheight]{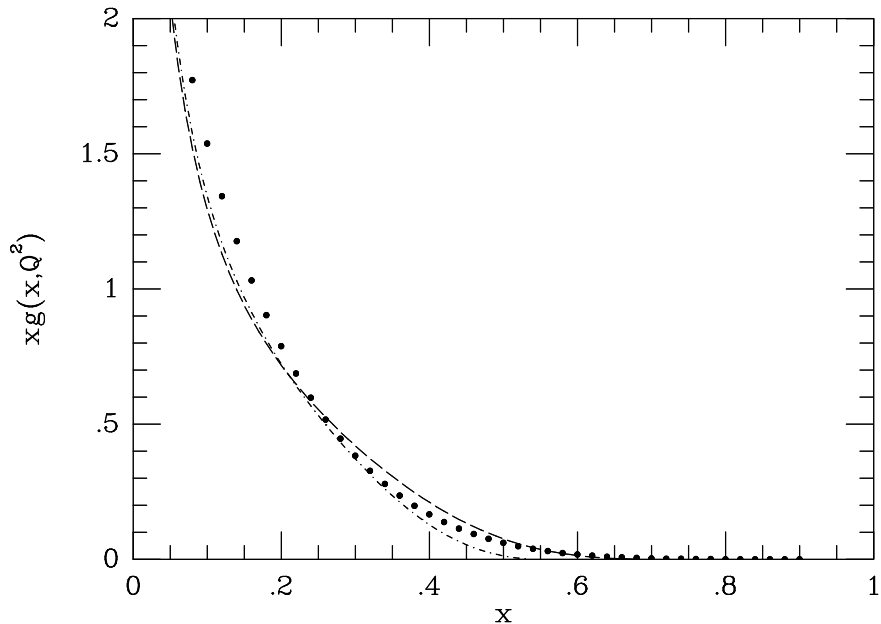}}
\end{figure}

In the polarized case, the spin distribution function for
the proton, which are too large for the model calculation,
the famous proton spin problem, decreases and approaches the
data via the same RG mechanism. In this way  the spin is
transferred to the new components and the problem greatly
disappears. In Fig. 3 the remaining discrepancy between the
model calculation and the data after evolution can be seen.

\begin{figure}
\caption{We show the spin structure function $g_1$ for the
proton. The dashed curve represents the results of a Quark
Model calculation evolved at NLO to the scale of the data
($10 GeV^2$). The full two curves represent the calculation
in the $ACMP$ scenario, within the same quark model,
for two parametrizations of the quark structure
functions. The data have been taken from \cite{datap}.}
\rotatebox{270}{\includegraphics[height=.5\textheight]{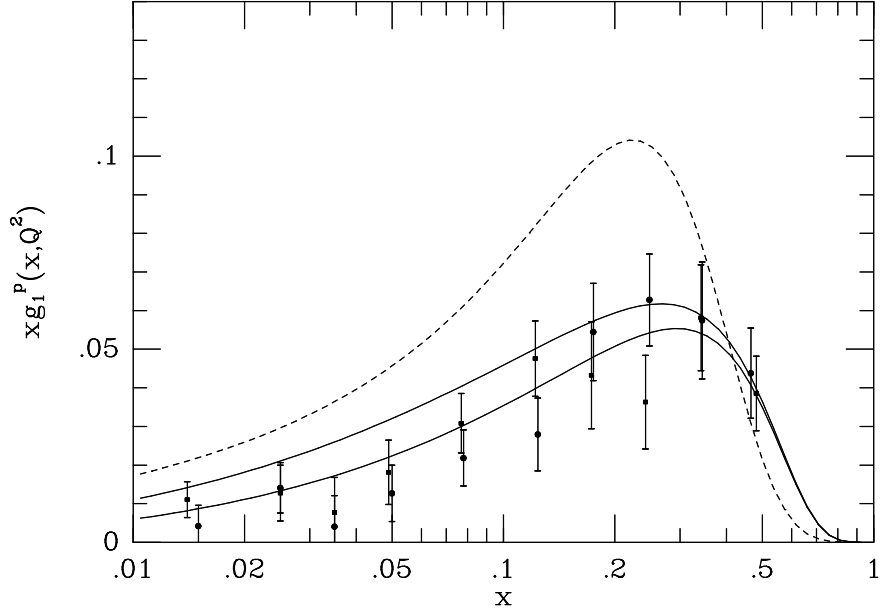}}
\end{figure}

If one aims at a quantitative agreement with the data, the
conventional low energy models have to be modified to
include,  higher momentum  and higher angular momentum
components for the quarks, and sea components at the
hadronic scale. Moreover the experimental gluon
distributions, at present extracted in a very indirect way,
if taken at face value, imply the need for soft gluons at
the hadronic scale. Moreover in the case of the spin parton
distribution, the anomaly contribution helps in the
explanation of the data.

{\bf 2) Transversity distribution \cite{scopetta2}}

The feasibility of measuring chiral-odd parton distribution
functions in polarized Drell-Yan and semi-inclusive
experiments has renewed theoretical  interest in their
study. Models of hadron structure have proven successful in
describing the gross features of the chiral-even structure
functions. Similar expectations motivated our study of the,
experimentally unknown,  transversity parton distributions
with these models. We confirmed, by performing a NLO
calculation,  the diverse low $x$  behaviors of the
transversity and spin structure  functions at the
experimental scale and showed that it is fundamentally a 
consequence of the different behaviors under evolution of
these functions.  The  inequalities of Soffer establish
constraints between data and model  calculations of the
chiral-odd transversity function. The approximate 
compatibility of our model calculations with these
constraints  confers credibility to our estimates.

{\bf 3) Skewed parton distributions}

A new type of observable, the so called skewed parton
distributions (SPD), have been intensively studied in the
last years \cite{xi}. The SPDs generalize and interpolate
between the ordinary parton distributions and the elastic
form factors and therefore contain rich structural
information. They have been instrumental in understanding
the Orbital Angular Momentum (OAM) and furthermore the Deeply 
Virtual Compton Scattering
(DVCS) process has been proposed as a practical way to
measure them. From the point of view of parton physics they
appear, similarly to the conventional distributions, as
light cone matrix elements of the quark-gluon operators,
however here the initial and final states have different
momenta, and in this way there is an additional
t-dependence besides the conventional $x$  dependence. A
model calculation within the MIT bag framework has provided
estimates about their magnitude which serve as guidance for
future experiments \cite{xi1}.

i) Orbital angular momentum \cite{scopetta1}

We have studied OAM twist-two
parton distributions, for the relativistic MIT bag model
and for non-relativistic potential quark models.  The
contribution of quarks OAM to the nucleon spin evolves at
high $Q^2$ to a vanishingly small value, while that of the
gluons increases dramatically.  As expected by general
arguments, the large gluon OAM contribution is almost
canceled by the gluon spin  contribution. At large $Q^2$
the gluons contribute  $50\%$ to the angular momentum and
the quarks carry only spin.

ii) Twist three contributions to DVCS \cite{anikin}

The study of the gauge invariance of the DVCS amplitude
leads to the inclusion of higher twist components
\cite{anikin1}.  We have performed an extensive study of
the DVCS amplitude within a bag model framework of the
single spin asymmetry in the case of spin 0 systems. Our
results imply that the choice of kinematics is crucial in
order to observe certain amplitudes and therefore unravel
the structure of the system.

\section{Towards an unified picture of constituent and
current quarks}

Our basic assumption has been that gluon and sea
bremsstrahlung are the source of difference between the
constituents and the current quarks. However the data seem
to indicate that the hadronic structure is more complex,
with primordial sea quarks (antiquarks) and gluons. Thus the 
analysis thus far implies that constituent models have to be of
greater complexity in order to describe,  simultaneously, low
and high energy data. Next, we analyze one way to generate
this complex
structure from a  quark model,
by assuming that constituent quarks are
non elementary and therefore have  partonic
structure. These ideas where investigated a long time ago
by two groups, Altarelli et al. \cite{acmp}, starting from
a quark model scenario, and Kuti and Weisskopf
\cite{kuti}, who defined a more complex scenario which
contained sea and gluons at the hadronic scale. We have
studied the consequences of the former approach.

\subsection{Current structure from the Constituent Quarks}

We have gone beyond the bremsstrahlung formalism by
incorporating structure to the constituent quarks following the
procedure we have called ACMP \cite{acmp}. Within this
approach constituent quarks are effective particles made up
of point-like partons (current quarks, antiquarks and
gluons), interacting by a residual interaction described by
a quark model \cite{scopetta}. The structure of the hadron is
obtained by a convolution of the constituent  quark model
wave function with the constituent quark structure
function. For a proton made up of $u$ and $d$ quarks,

\begin{equation} 
f(x,\mu_0^2) = \int^1_x
\frac{dz}{z}[u_0(z,\mu_0^2)\Phi_{uf}(\frac{x}{z},\mu_0^2) +
d_0(z,\mu_0^2)\Phi_{df}(\frac{x}{z},\mu_0^2)], 
\label{ccq}
\end{equation}
where $\mu_0^2$ is the hadronic scale, $f = q_v, q_s, g$ (valence 
quarks, sea quarks and gluons respectively) and
$\Phi$ represents the constituents probability in each
quark and has been parametrized following general arguments
of $QCD$ as 

\begin{equation} \Phi_{qf}(x,\mu_0^2) = C_f
x^{a_f}(1-x)^{A_f-1}.   
\end{equation}  

The constants have been fixed by Regge phenomenology and
the choice of the hadronic scale ($\mu_0 = 0.34$ GeV$^2$).

The discussion can be generalized to the polarized
structure functions \cite{scopettas}. The procedure is
able to reproduce the data extremely well and in this
framework the so called spin problem does not arise.

\subsection{Applications}

{\bf 1) Unpolarized parton distributions \cite{scopetta}}

Using  that the constituent quark is a composite system of
point-like partons, we construct the  parton distributions
by a convolution between constituent quark momentum
distributions and constituent quark structure functions, 
Eq.(\ref{ccq}). 

The different types and functional forms of the structure
functions of the  constituent quarks, $\Phi$, are derived from three
very natural assumptions: 

\begin{itemize} 
\item[  i)]The point-like partons are determined by $QCD$,
therefore, they are quarks, antiquarks and gluons; 

\item[ ii)]
Regge behavior for $x\rightarrow 0$ and duality ideas;

\item[iii)] invariance under charge conjugation and
isospin. 

\end{itemize}

These considerations define in the case of the valence
quarks the following structure function,

\begin{equation}
\phi_{qq_v}({x},Q^2)
= { \Gamma(A + {1 \over 2}) \over 
\Gamma({1 \over 2}) \Gamma(A) }
{ (1-x)^{A-1} \over \sqrt{x} }.
\label{csf1}
\end{equation}

For the sea quarks the corresponding structure function
becomes,

\begin{equation}
\phi_{qq_s}({x},Q^2)
= { C \over x } (1-x)^{D-1},\label{csf2}
\end{equation}
and in the case of the gluons we take

\begin{equation}
\phi_{qg}({x},Q^2)
= { G \over x } (1-x)^{B-1}~.\label{csf3}
\end{equation}

The other ingredients of the formalism, i.e., 
the probability distributions for 
each constituent quark, are defined according to the
procedure of Traini
et al. \cite{Traini}, that is, a constituent quark has
a probability
distribution determined by Eq.(\ref{traini}).

Our last assumption relates to the scale at which the
constituent quark  structure is defined. We choose for it
the  hadronic scale $\mu_0^2$. This hypothesis fixes $all$
the parameters of the approach except one, which is fixed
by looking at the low $x$ behavior of the $F_2$ structure
function at the hadronic scale, where the sea in known to
be dominant.

The resulting parton distributions and structure functions 
are evolved to the experimental scale and good agreement 
with the available DIS data is achieved (See Fig. 1). In
Fig. 2 we show the gluonic components generated in the ACMP
scheme for two models. The primordial sea and gluon
components at the hadronic scale are instrumental in
achieving a good agreement with the experimental
observation.

When compared with a similar calculation using
non-composite constituent quarks, the accord with
experiment of the present calculation becomes  impressive.
We therefore conclude that DIS data are consistent with a
low energy scenario dominated by composite,  mainly
non-relativistic constituents of the nucleon.

{\bf 2) Polarized parton distributions \cite{scopettas}}

The previous discussion can be generalized to the polarized
case. The functions
$\Phi_{ab}$ now specify spin and flavor. 

Let

\begin{equation}
\Delta q (x,\mu_0^2) = q_+ (x,\mu_0^2) - q_- (x,\mu_0^2)
\end{equation}
where $\pm$ label the quark spin projections and $q$
represents any flavor. 
The generalized $ACMP$ approach implies 
\begin{equation}
q_i(x,\mu_0^2) = \int_x^1 \frac{dz}{z} \sum_j
(u_{0j}(z,\mu_0^2) 
\Phi_{u_{j}q_i} (\frac{x}{z},\mu_0^2) + d_{0j}(z,\mu_0^2)
\Phi_{d_{j} q_j}(\frac{x}{z},\mu_0^2))
\end{equation}
where $i=\pm$ labels the partonic spin projections and
$j=\pm$ the constituent
quark spins. Using spin symmetry we arrive at \footnote{We
omit writing
explicitly the hadronic scale dependence from now on.}

\begin{equation}
\Delta q (x) =\int_x^1 \frac{dz}{z} ( \Delta u_0 (z) \Delta
\Phi_{uq}(\frac{x}{z}) + \Delta d_0 (z)\Delta
\Phi_{dq}(\frac{x}{z}))
\end{equation}
where $\Delta q_0 = q_{0+} - q_{0-}$, and 

\begin{equation}
\Delta \Phi_{uq} = \Phi_{u+q+} - \Phi_{u+q-}
\end{equation}
\begin{equation}
\Delta \Phi_{dq} = \Phi_{d+q+} - \Phi_{d+q-}
\end{equation}

We next reformulate the description in term of the
conventional valence and sea quark separation, i.e.,

$$\Delta q (x) = \Delta q_v (x) +\Delta q_s (x) $$

After a series of simplifying assumptions we obtain for the
various polarized parton distributions the following
expressions:

\begin{equation}
\Delta q_v (x) =\int_x^1 \frac{dz}{z} \Delta q_0 (z) \Delta
\Phi_{qq_v} 
(\frac{x}{z}),
\end{equation}
for the valence quarks,

\begin{equation}
\Delta q_s (x) = \int_x^1 \frac{dz}{z} (\Delta u_0 (z)  +
\Delta d_0 (z)) 
\Delta \Phi_{qq_s}(\frac{x}{z}),
\label{deltasea}
\end{equation}
for the sea quarks, and

\begin{equation}
\Delta g (x) = \int_x^1 \frac{dz}{z} (\Delta u_0 (z) +
\Delta d_0 (z)) 
\Delta \Phi_{q g}(\frac{x}{z})
\end{equation}
for the gluons

Thus the  $ACMP$ procedure can be extended to the polarized case
just by introducing three additional structure functions
for the constituent quarks: $\Delta \Phi_{q q_v}$, $\Delta
\Phi_{q q_s}$ and $\Delta \Phi_{q g}$.

In order to determine the polarized constituent structure
functions we add some assumptions which will tie up the
constituent structure functions for the polarized and
unpolarized cases completely, reducing dramatically the
number of parameters. They are:

\begin{itemize}

\item [iv)] factorization assumption: $\Delta \Phi$ cannot
depend upon the quark 
model used, i.e, cannot depend upon the particular $\Delta
q_0$;

\item [v)] positivity assumption: the positivity constraint
$\Delta \Phi \leq
\Phi $ is saturated for $x = 1$.

\end{itemize}

These additional assumptions determine completely the
parameters of the polarized constituent structure functions.

Using unpolarized data to fix the parameters we achieve
good agreement with the polarization experiments for the
proton (see Fig. 3), while not so for the neutron. By relaxing our 
assumptions for  the sea distributions,   we define new
quark functions for  the polarized case, which  reproduce
well the proton data and are in better agreement with the 
neutron data (see discussion in ref. \cite{scopettas}).    

When our results are compared with similar calculations
using non-composite  constituent quarks the accord with the
experiments of the present scheme  is impressive. We
conclude that, also in the polarized case, DIS data are
consistent with a low energy scenario dominated by
composite constituents of the nucleon.

\section{Concluding remarks}

The high energy parton distributions when evolved to a low
energy scale appear to indicate that a valence picture of
hadron    structure arises. This valence picture is well
represented theoretically by Quark Models which are very
successful in explaining the low energy properties of
hadrons. We have developed a formalism based on a
laboratory partonic description which connects the parton
distributions with the momentum distributions of the
constituents giving us a description of partons in terms of
Quark Model wave functions. Our basic assumption is that
gluon and sea bremsstrahlung are the source of difference
between the various momentum scales. We have implemented
the Renormalization Group program by defining a hadronic
scale and using as initial, non perturbative, conditions
those obtained from the parton distributions of the low
energy model.

Our analysis, based on a NLO formalism of evolution, has
shown that the perturbative scheme is applicable to the low
energy scales of interest. The formalism used has the
correct support for the parton distributions and allows the
discussion of a large class of Quark Models.

The results of
our calculations show that low energy models, with their
parameters fixed by low energy properties, tend to give a
qualitative description of the data. Fig. 1 is very
clarifying in this respect. This feature allows
one to use them in order to be predictive in new
observations.

The next step, which our formalism allows, is to proceed to
define models which describe quantitatively the data at all
energy scales. Our analysis has shown that present models
are too naive. The new models seem to require: primordial
gluons and sea.

The limitations associated with naive Quark Models of DIS
data can be overcome by a very appealing scheme where the
constituent quarks are not elementary. Partons (the
quarks,  antiquarks and gluons of $QCD$) at the hadronic
scale are generated by  unveiling the structure of the
constituent quarks. We have seen that  incorporating this
structure in a very physical way improves notably  the
agreement with the DIS data (See Fig. 1). From the point of view of the
calculation, we must stress, that no parameters of the
model have been changed with respect to the original  fit
to the low energy properties. The new parameters arising
from the description of the constituent quark structure
functions have been adjusted to describe the input scenario
according to the hadronic scale philosophy. In this way
the sea and gluon distributions are generated in a
consistent way (see Fig. 2).

The same analysis can be easily performed for the polarized
case. Using a physically motivated minimal prescription for
the polarized case,  with no additional parameters, we are
able to obtain a good prediction of the the proton data (see
Fig. 3).
The minimal procedure fails, however,  to reproduce  the
accurate neutron data. Relaxing the minimal procedure, with
the addition of only one new parameter to define the
polarized sea, we obtain a  significantly improved
description also for the neutron data \cite{scopettas}.
The calculation has also clarified the role of the gluons
and the valence quarks. It is clear that the gluons become
important through the evolution process, i.e., it is the
soft bremsstrahlung gluons which acquire a large portion of
the partonic spin.

We would like to stress that within our procedure the {\it
spin problem}, as initially presented,  does not arise. 
The constituent quarks carry all of the polarization. When
their structure is unveiled this polarization is split
among their different partonic contributions  in the manner
we have described and which is consistent with the data.
The quality of both unpolarized and polarized data thus 
far analyzed confirm the validity of the approach. We have
showed also, that with very reasonable assumptions, the
scheme becomes highly predictive, a feature which is
necessary for the planning of future experiments.

We feel safe to conclude that, the current quarks seen at
the parton level  seem to be embedded in the composite
constituent quarks seen at lower $Q^2$.  An unified picture
of current quarks, successfully describing DIS,   and
constituent quarks, successfully describing static
properties is  possible.

\begin{theacknowledgments}

Sergio Scopetta has been a main contributor to much of the
work presented here. It has been a great pleasure to work
with him. My collaboration with Marco Traini comprises a
long period of time. During all these years it has been
very inspiring to interact with him. I acknowledge
conversations with I.V. Anikin, D. Binosi, R. Medrano and
S. Noguera on Skewed Parton Distributions and gauge
invariance. I thank the organizers for the invitation to
present my work in this most beautiful city, and in
particular Jiri Adam, for his continuous help and assistance
before and during the conference. This work was supported
in part by DGICYT grant PB97-1227 and La Generalitat
Valenciana with  a travel grant.

\end{theacknowledgments}

% This is the version of References if BiBTeX is used
% Compliant BiBTeX styles are aipproc  (for use with natbib) 
%                         and aipprocl (if natbib is missing at the site).
%\bibliography{sample}

\begin{thebibliography}{99} 

\bibitem{Gell-Mann} Kokkedee, J.J.J., {\it The Quark Model},
W. A. Benjamin, New York, 1969; Lichtenberg, D.B., {\it Unitary Symmetry and
Elementary Particles}, Academic Press, Inc. New York, 1978. 

\bibitem{Feynman} Feynman, R.P.,{\it Photon-Hadron
Interaction}, W. A. Benjamin, New York, 1972.
\bibitem{QCD}  Muta, T., {\it Foundations of Quantum
Chromodynamics}, World Scientific, Singapore 1987 ; 
Field, R.D., {\it Applications of perturbative $QCD$}, Addison
Wesley Pub. Co., New York, 1989; Yndurain, F.J., {\it The Theory of
the Quark and Gluon Interactions}, Springer Verlag, Heidelberg, 1999.

\bibitem{RAL94} Martin, A.D.,  Stirling W.J., and 
Roberts, R.G., Ral Report 94-055; ibid Ral Report 95-021.

\bibitem{GRV} Gl\"uck, M., and Reya, E., {\it  Phys. Rev.} {\bf D14}
 3024 (1976); Reya, E., {\it Phys. Rep. } {\bf 69}  195 (1981); 
Gl\"uck, M., Reya, E., and  Vogt, A., {\it Z. Phys.} {\bf C48} 471 (1990); 
 Gl\"uck, M., Reya, E., and Vogt, A., {\it Z. Phys.} {\bf C53}
 127 (1992); {\it Z. Phys. }{\bf C67}  433 (1995).

\bibitem{Parisi}  Parisi, G., and  Petronzio, R., {\it Phys. Lett.}
{\bf B62}  331 (1976).

\bibitem{models} Alvarez-Estrada, R.F.,  Fern\'andez, F., 
S\'anchez G\'omez, J.L., and Vento, V., "Model of Hadron
Structure based on QCD", {\it Lecture Notes in Physics {\bf 259}}, 
Springer Verlag, Heidelberg, 1986. 

\bibitem{Jaffe} Jaffe, R.L., and  Ross, G.C., {\it Phys.
Lett. } {\bf B93} 313 (1980).  


\bibitem{Traini} Traini, M., Zambarda, A., and Vento, V., {\it Mod.
Phys. Lett.}  {\bf 10}  1235 (1995); Ropele, M., Traini, M.,
and Vento, V., {\it Nucl. Phys.} {\bf A584}  634 (1995); Traini, M.,
Vento, V.,  Mair, A., and Zambarda, A., {\it  Nucl. Phys.} {\bf A614} 472
(1997). 

\bibitem{betz}  Jaffe, R.L., {\it Phys. Rev.} {\bf D11} 1953 (1975);  
 Betz, M., and Goldflam, R., {\it Phys. Rev.} {\bf D28} 2848 (1983). 


\bibitem{acmp} Altarelli, G., Cabibbo, N., Maiani, L., and 
Petronzio, R., {\it Nucl. Phys.} {\bf B 69}  531 (1974).


\bibitem{scopetta} Scopetta, S.,  Vento, V.,  and Traini, M.,
{\it Phys. Lett.} {\bf B412}  64 (1998).

\bibitem{scopetta1}Scopetta, S., and  Vento, V., {\it Phys. Lett.}
{\bf B460}  8 (1999), {\it Phys. Lett.}  {\bf B474}  235 (2000). 

\bibitem{scopetta2} Scopetta, S., and Vento, V., {\it Phys. Lett.} {\bf
B424} 25 (1998).

\bibitem{bil} Bijker, R., Iachello, F., and  Leviatan, A., {\it  Ann.
Phys.} {\bf 236}  69 (1994) ; {\it Phys. Rev.} {\bf C54} 1935 (1996) ;
{\it Phys. Rev. } {\bf D55}  2862 (1997).



\bibitem{xi} Radyuskin, A.V., {\it Phys. Lett.} {\b 380} 417 (1996)
; {\it Phys. Lett.} {\bf 385} 333 (1996) ; Ji, X., {\it Phys. Rev. Lett.}
{\bf 78}   610 (1997); {\it Phys. Rev.} {\bf D55}  7114 (1997).

\bibitem{xi1}
Ji, X.,  Melnitchouk, W.,  and Song, X., {\it Phys. Rev.} {\bf D56}
 5511 (1997).



\bibitem{anikin}  Anikin, I.V., Binosi, D., Medrano, R., 
Noguera, S., and Vento, V.,  work in preparation.


\bibitem{anikin1}
Anikin, I.V., Pire, B.,  and Teryaev, O.V.,
{\it Phys. Rev.} {\bf D62}  071501 (2000).


\bibitem{kuti} Kuti, J., and  Weiskopf, V.F., {\it Phys. Rev.} {\bf
D11}  3418 (1974).



\bibitem{scopettas}  Scopetta, S., Vento, V., and  Traini, M.,
{\it Phys. Lett.} {\bf 442}  28 (1998). 



\bibitem{dataup}
 Lai, H.L. et al., {\it Phys. Rev.} {\bf D51}  4763 (1995).

\bibitem{ik} Isgur, N., and Karl, G., {\it Phys. Rev.} {\bf D18} 4187 
(1978) ; {\bf D19}  2653 (1979) ; {\bf D23}  817 (1981)  (E).


\bibitem{datap}
EMC Collaboration,  Ashman, J., et al., {\it Nucl. Phys.} {\bf B328} 1
(1989);
SMC Collaboration,  Adams, D., et al., {\it Phys. Rev.} {\bf D56} 5330
(1997).

\end{thebibliography}

% Alternatively, the references in the standard LaTeX method
% should follow these conventions:

\end{document}